\documentclass[superscriptaddress,twocolumn,showpacs,preprintnumbers,amsmath,
amssymb,10pt,aps,prb]{revtex4-1}
\usepackage{graphicx}
\usepackage{color}
\usepackage{ulem}

\makeatletter
\newcommand\colorsout[1]{\bgroup \markoverwith{\textcolor{#1}{\rule[0.5ex]{2pt}{0.4pt}}}\ULon}

\begin{document}
\title{Structural and magnetic properties of FeMn$_x$ ($x=$1...6) chains supported on Cu$_2$N / Cu (100)}
\author{Deung-Jang Choi}
\affiliation{CIC nanoGUNE, Tolosa Hiribidea 78, 20018 Donostia-San Sebastian, Spain}
\author{Roberto Robles}
\affiliation{Catalan Institute of Nanoscience and Nanotechnology (ICN2), CSIC and The Barcelona Institute of Science and Technology, Campus UAB, Bellaterra, 08193 Barcelona, Spain}
\author{Jean-Pierre Gauyacq}
\affiliation{Institut des Sciences Mol\'eculaires d'Orsay, ISMO, CNRS, Universit\'e Paris-Sud,  Universit\'e Paris-Saclay, B\^atiment 351, 91405 Orsay Cedex, France}
\author{Markus Ternes}
\affiliation{Max Planck Institute for Solid State Research, 70569 Stuttgart, Germany}
\author{Sebastian Loth}
\affiliation{Max Planck Institute for the Structure and Dynamics of Matter, 22761 Hamburg, Germany}
\affiliation{Max Planck Institute for Solid State Research, 70569 Stuttgart, Germany}
\author{Nicol\'as Lorente}
\affiliation{Centro de Fisica de Materiales, CFM/MPC (CSIC-UPV/EHU), Paseo Manuel de Lardizabal 5, 20018 Donostia-San Sebastian, Spain}
\affiliation{Donostia International Physics Center (DIPC), Paseo Manuel de Lardizabal 4, 20018 Donostia-San Sebastian, Spain}

\begin{abstract}
Heterogeneous atomic magnetic chains are built by atom manipulation on
a Cu$_2$N/Cu (100) substrate. Their magnetic properties
are studied and rationalized by a combined scanning tunneling
microscopy (STM) and density functional theory (DFT) work
completed by model Hamiltonian studies. The chains are built using Fe
and Mn atoms ontop of the Cu atoms along the N rows of the Cu$_2$N surface.  Here, we
present results for FeMn$_x$ ($x$=1...6) chains emphasizing the
evolution of the geometrical, electronic, and magnetic properties with
chain size. By fitting our results to a Heisenberg Hamiltonian we have
studied the exchange-coupling matrix elements $J$ for different chains. For
the shorter chains, $x \leq 2$, we have included spin-orbit effects in the
DFT calculations, extracting the magnetic anisotropy energy. Our results 
are also fitted to a simple anisotropic spin Hamiltonian and we have extracted
values for the longitudinal-anisotropy $D$ and transversal-anisotropy $E$
constants. These parameters together with the values for $J$ allow us
to compute the magnetic excitation energies of the system and
to compare them with the experimental data.
\end{abstract}

\date{\today}

\maketitle

\section{Introduction}

Bottom-up approaches give access to
systems of very reduced dimensionality with
unique physical properties. Among these systems,
{chains of a single-atom cross section} are of great interest.~\cite{wang_2011}
When the chains are formed by magnetic atoms, spin-spin
correlations can come into place leading to new
phenomena and applications.~\cite{brune_2006} Building
structures from their atomic constituents can be achieved
by the atomic-manipulation capabilities of 
the scanning tunneling
microscope (STM). Magnetic atoms have been positioned one by one
at different distances and with different arrangements on a variety of
substrates.\cite{hirjibehedin_2006,khajetoorians_2012,loth_2012,holzberger_2013,yan_2014}
Spin-polarized STM~\cite{wiesendanger} or inelastic
electron tunneling spectra (IETS)~\cite{heinrich_2004,PSS}
have granted us a detailed vision of the magnetic mechanisms
at play on the atomic scale.

Recent developments in density functional
theory (DFT) have also permitted us to
attain a deep understanding
of the phenomena revealed by the above experiments.
Due to their experimental interest,
Mn chains are amongst the most studied ones.\cite{lounis_2008,rudenko_2009,lin_2011,urdaniz_2012,Tao2015,choi_2016}
Fe chains are also well known both experimentally
and theoretically.\cite{khajetoorians_2012,loth_2012,nicklas_2011,gauyacq_2013,khajetoorians_2013,spinelli_2014,spinelli_2015,Tao2015}.
The work by
 Lin and Jones~\cite{hirjibehedin_2007,lin_2011} 
on Fe, Co and Mn atoms 
on Cu$_2$N/Cu (100) 
reveals that atomic spins maintain their
nominal values on the surface ($S_{Fe}$= 2, $S_{Mn}$=5/2 and $S_{Co}$=
3/2), showing the interest of using Cu$_2$N
to preserve much of the magnetic identity of transition-metal (TM) atoms.
Studies of Fe chains~\cite{nicklas_2011} 
and of Mn~\cite{rudenko_2009} reveals that close-packed
TM chains on Cu$_2$N/Cu (100) couple
antiferromagnetically due to a N-mediated superexchange
mechanism,
largely explaining experimental findings.\cite{hirjibehedin_2007,loth_2012}

In the present work, we report on a different type of magnetic atomic
chain. These chains are heterogeneous, including two types of magnetic
atoms, Fe and Mn, on a Cu$_2$N/Cu (100) substrate. An initial account of
the results has been given in a separate publication.~\cite{choi_2014}
The experimental study is based on STM and IETS results of FeMn$_x$
($x=1,\cdots, 6$) and confirms that, as in the previous cases, the magnetic
ordering along the chain is antiferromagnetic. The different anisotropy
of Fe and Mn on Cu$_2$N/Cu (100) leads to two possible orientations of
the magnetic moments (along the chain and out-of-plane).  Contrary to
the case of homogeneous chains, this may indicate a non-collinear
arrangement. When $x$ is an odd number, a simple-minded evaluation of
the total spin yields 1/2, which is compatible with the appearance of a
Kondo feature at zero bias.~\cite{choi_2014} However, when $x$ is even,
the expected spin is 2 and, correspondingly, no Kondo peak is observed.
In the present work, we perform extensive calculations and compare them
with the experiments. The comparison permits us to conclude on the spin
arrangement (it is collinear along the chain) and on the behavior of
the exchange coupling between the two different atomic species of the
chain in good agreement with the experimental observations. The 
antiferromagnetic coupling is confirmed and traced back to the
N-mediated superexchange mechanism as in the homogeneous chains. Finally,
the obtention of model Hamiltonians to study the magnetic structure of
these chains is discussed within the framework of
DFT calculations.~\cite{gauyacq_2013,kepenekian_2014}

\section{Experimental method}

Experiments were performed in an ultrahigh vacuum low temperature STM at
a base temperature of 0.5 K as has been partially reported 
in Ref.~[\onlinecite{choi_2014}]. 
Differential conductance was used as a local
spectroscopic tool that gives information via inelastic
electron tunneling (IETS).\cite{heinrich_2004,PSS}
The differential conductance was directly
measured using lock-in detection with 72-$\mu$V rms modulation at ~691 Hz
to the sample bias voltage V. 

The Cu (100) surface was cleaned by Ar sputtering
and then annealed up to 850 K. After having big terraces of Cu(100)
crystal, a monolayer of Cu$_2$N was formed as a decoupling layer by 
nitrogen sputtering and post-annealing at 600 K. Single
Fe and Mn atoms were deposited onto the pre-cooled surface. Pt-Ir tips were
prepared by sputter-anneal cycles and coated with copper {\it in vacuo} by
soft indentations into the Cu bridges. The tip status was monitored through
STM images and controlled to manipulate the atoms. 
All atomic chains were built using vertical atomic manipulation. 
After identifying a given adatom by its spectroscopic features, we picked it up by voltage pulsing, dropped it off on a nitrogen site 
and hop it to a copper site. We build the chain
 atom by atom in
 a close-packed configuration to ensure an AF coupled spin chain. 
By doing so, all different kinds of FeMn$_x$ ($x$=1... 10) chains were constructed.

\begin{figure}[ht]
\includegraphics[width=0.7\columnwidth]{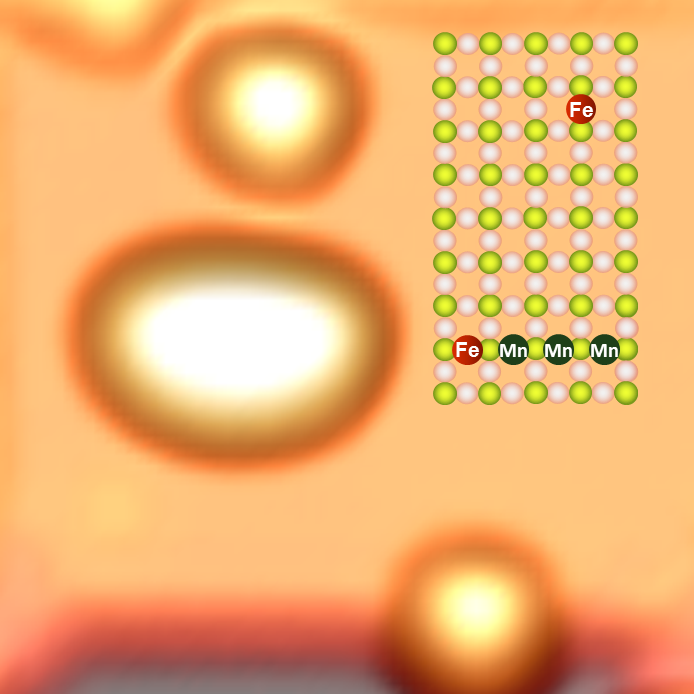}
\caption{\label{stm} {Constant current STM image of a FeMn$_3$
chain assembled on a Cu$_2$N layer  and 
atomic scheme (on the right of the figure) that can be inferred from 
the position of the upper-right Fe atom and manipulation
with the STM tip. The STM image was scanned at 10 mV and
100 pA. The scanned area is 6 $\times$ 6 nm$^2$. The transition metal atoms 
were always placed ontop of a Cu atom between two nitrogen atoms.
The topography was processed with WSXM\cite{WSXM}.}
} 
\end{figure} 

\section{Theoretical method}
\label{theory}

{\it Ab initio} calculations are performed within the
density-functional theory (DFT) framework as implemented in
the VASP code.\cite{kresse_efficiency_1996} We expand the wave
functions using a plane-wave basis set with a cutoff energy of
300~eV. Core electrons are treated within the projector augmented wave
method.\cite{bloechl_projector_1994, kresse_ultrasoft_1999} The PBE form
of the generalized gradient approximation is used for the exchange and
correlation functional.\cite{perdew_generalized_1996}

\begin{figure}[ht]
\includegraphics[width=1.0\columnwidth]{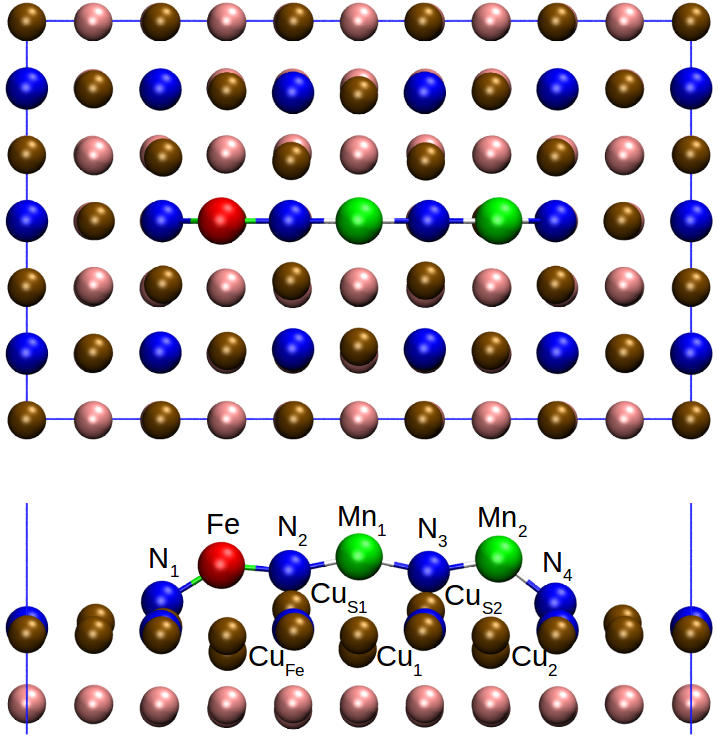}
\caption{\label{trimer} Top (upper panel) and side (down panel) views of
the relaxed geometry of FeMn$_2$. We show Fe atoms in red; Mn atoms in
green; N atoms in blue; Cu atoms in the first layer in brown; and the
rest of Cu atoms in pink. Blue lines delimit the unit cell used in the
calculations. The vacuum side of the unit cell and the three bottom Cu layers
have been removed for pictorial reasons.} 
\end{figure}

The system consists of a Cu(100) surface covered by a Cu$_2$N layer.
To model the surface we use a slab geometry with four Cu layers plus
the Cu$_2$N layer.  Following the above experimental procedure, Fe and
Mn atoms are deposited on top Cu atoms, forming a chain in the [010]
direction. We use a unit cell {[3 $\times$ ($x$+3)] in units of the
bulk lattice parameter such that the length along the [010] direction
increases as  ($x$+3)}, where $x$ is the number of Mn atoms. In this way,
we keep the distance between chain images constant for all sizes, being
of 3 lattice constants in the unrelaxed configuration. The unit cell
for FeMn$_2$ is shown in Fig.~\ref{trimer} as an example.  The bottom
Cu layer was kept fixed and the remaining atoms were allowed to relax
until forces were smaller than 0.01~eV/\AA. The $k$-point sample was
varied accordingly to the unit cell, and tests were performed to assure
its convergence.  The charge and magnetic moments have been calculated
using the Bader analysis.~\cite{tang_bader_2009}

A critical aspect of the calculations is the use of
a static Coulomb charging energy $U$. Lin and Jones~\cite{lin_2011}
perform a constrained DFT calculation to evaluate $U$. For
Mn they find $U$= 4.9 eV when Mn is sitting ontop a surface Cu
atom, while it is reduced to $U$= 3.9 eV ontop a N site. This is in agreement
with their previous result~\cite{hirjibehedin_2007}, where they found $U$= 5.0 eV
for Mn ontop of Cu, and $U$= 2.0 eV for Fe. To the best of our knowledge, 
these are the only actual computations of $U$ for Mn on the Cu$_2$N/Cu (100)
substrate. Rudenko {\it et al.}\cite{rudenko_2009} take an
effective value $U-J$=5 eV {where $J$ is the intra-atomic
exchange coupling.}  Nicklas {\it et al.}~\cite{nicklas_2011}
do not take any value of $U$ for Fe. As Lin and Jones~\cite{lin_2011}
show, the actual value of $U$ greatly affects the computed
exchange couplings, indeed they find a difference of a factor
of 2 between their calculations with $U$= 4.9 eV and without $U$,
finding DFT+U results in better agreement with the experiment.

In this work, we have used 
the GGA+U method of Dudarev~\cite{dudarev_1998} 
with a $U_{\texttt{eff}}=U-J$ of 1~eV for Fe and 4~eV for Mn.
The chosen values correspond to roughly subtracting $J\approx 1$ to
$U=4.9$ eV ($U=2$ eV) as computed by Lin and Jones~\cite{lin_2011} for Mn (Fe) ontop a Cu atom. 
As we will see, the values of $U_{\texttt{eff}}$
become critical for the determination of the exchange couplings. 
The effect of different values for $U_{\texttt{eff}}$ is discussed
below. 

To rationalize our results we have fitted them to a Heisenberg Hamiltonian in the form:
\begin{equation}
\label{heisenberg}
H=-\sum_{i,j>i} J_{ij}{S}_i \cdot{} {S}_j,
\end{equation}
where $J_{ij}$ are the exchange couplings between spins ${S}_i$ and ${S}_j$.
Evaluating different magnetic configurations  we have been able to extract
the $J_{ij}$ values by fitting the DFT energies to the Heisenberg Hamiltonian, Eq.~(\ref{heisenberg}).

For the shorter chains (FeMn and FeMn$_2$) we have included spin-orbit
coupling (SOC)~\cite{vasp_manual,hobbs_2000} into our scalar-relativistic
Hamiltonian. We have considered a simple anisotropic spin
Hamiltonian, in the form
\begin{equation}
\label{ani}
H=DS_z^2+E(S_x^2-S_y^2).
\end{equation}
Here $x,y$ and $z$ are such that $z$ is the computed easy axis, while $x$ and $y$
are two orthogonal directions in the plane perpendicular to $z$
(hard-plane) {in principle, different from the surface directions}. The energy of the system has been self-consistently
calculated including SOC for the different orientations $x$,
$y$, and $z$ of the magnetization axis and, from these energies,
values for $D$ and $E$ in Eq.~(\ref{ani}) have been fitted using
the evaluated magnetic moments.  
 
\begin{table*}[t]
\centering
\begin{ruledtabular}
\begin{tabular}{lcccccc}
 & d[Fe-Mn$_1$] & d[Mn$_1$-Mn$_2$] & d[Mn$_2$-Mn$_3$] & d[Mn$_3$-Mn$_4$] & d[Mn$_4$-Mn$_5$] & d[Mn$_5$-Mn$_6$]\\
        \hline
        FeMn       & 3.81 &      &  &  &  & \\
        FeMn$_2$   & 3.77 & 3.84 &  &  &  & \\
        FeMn$_3$   & 3.75 & 3.80 & 3.82 &  &  & \\
        FeMn$_4$   & 3.74 & 3.77 & 3.78 & 3.79 &  & \\
        FeMn$_5$   & 3.73 & 3.75 & 3.76 & 3.77 & 3.77 & \\
        FeMn$_6$   & 3.73 & 3.74 & 3.75 & 3.74 & 3.77 & 3.74 \\
\end{tabular}
\end{ruledtabular}
\caption[] {\label{distance1} Distances between the
different TM atoms for the 6 considered chains, (in \AA{}). Note that
the unrelaxed value would be 3.65~\AA~in agreement with the
DFT lattice parameter of the Cu substrate.}
\end{table*}

\section{Results} 

\subsection{Geometries and energetics}

Constant current images were obtained between atom manipulation sequences.
This permitted us to have  precise knowledge of the atomic arrangement. 
Unfortunately, the structure of the tip that is optimal for atom manipulation
is not necessarily good for image production and the obtained images do
not contain much information, as can be seen in 
Fig.~\ref{stm}. In the inset, we show the configuration
as inferred from the atomic manipulation procedure that corresponds
to a FeMn$_3$ chain.

The calculation yields precise insight on the actual geometries. In
Fig.~\ref{trimer} we show the FeMn$_2$ geometry as an example. As we
can see in the side view, after deposition of the TM
atoms the final chain includes the N atoms in between them. These atoms
are lifted from the surface. N atoms at the edge of the chain, like
N$_1$, are moved upwards by $\approx$0.7~\AA, while N atoms in between
TM atoms, like N$_2$, are lifted by $\approx$1.6~\AA{} with respect
to their unrelaxed positions in the bare CuN$_2$ surface. Therefore,
we can assume that the final chains have the form FeMn$_x$N$_{x+2}$. Of
course, this is just a conventional choice to identify the final chain,
since other atoms in the surface are also significantly disturbed and
could be considered as parts of the chain, like Cu$_{\text{S}1}$ and Cu$_{\text{S}2}$.

It is interesting to study the evolution of the geometry of the chains
as their size increases. In Table~\ref{distance1} we show the distance
between TM atoms within the chain for all considered sizes. We observe
that these distances are bigger than the unrelaxed value (3.65~\AA\ corresponding
to the PBE lattice parameter of Cu due to the arrangement of the chain on the surface). For
example, d[Fe-Mn$_1$] for FeMn is 3.81~\AA{}, 4\% bigger than the
unrelaxed value. This distance tends to decrease as the chain size is
increased, reaching 3.73~\AA{} for FeMn$_6$. %(2.5\%).
The same behavior can be observed for other positions in the chain. 
Another interesting distance to track involves the Cu atoms just
underneath Fe and Mn atoms. As can be seen in the side view of
Fig.~\ref{trimer}, these Cu atoms (Cu$_{\text{Fe}}$, Cu$_1$, Cu$_2$)
are pushed downwards by around 0.4~\AA{}. 
%\begin{table*}[t]
%\centering
%\begin{ruledtabular}
%\begin{tabular}{lccccccc}
% & d[Fe-Cu$_{\texttt{Fe}}$] & d[Mn$_1$-Cu$_1$] & d[Mn$_2$-Cu$_2$] & d[Mn$_3$-Cu$_3$] & d[Mn$_4$-Cu$_4$]
% & d[Mn$_5$-Cu$_5$] & d[Mn$_6$-Cu$_6$] \\
%        \hline
%        Fe         & 2.38 &  &  &  &  &  & \\
%        Mn         &      & 2.47 &  &  &  &  & \\
%        FeMn       & 2.39 & 2.51 &  &  &  &  & \\
%        FeMn$_2$   & 2.40 & 2.55 & 2.53 &  &  &  & \\
%        FeMn$_3$   & 2.41 & 2.56 & 2.54 & 2.54 &  &  & \\
%        FeMn$_4$   & 2.41 & 2.56 & 2.54 & 2.54 & 2.55 &  & \\
%        FeMn$_5$   & 2.42 & 2.56 & 2.54 & 2.55 & 2.54 & 2.55 & \\
%        FeMn$_6$   & 2.42 & 2.57 & 2.55 & 2.55 & 2.55 & 2.54 & 2.57 \\
%\end{tabular}
%\end{ruledtabular}
%\caption[] {\label{distance2} Distances (in \AA{}) between TM atoms and Cu atoms below them. Values
%for isolated Fe and Mn atoms on the surface are included for comparison.}
%\end{table*}
%We show the distances in
%Tab.~\ref{distance2} for all the calculated chains, and for isolated
%Fe and Mn atoms adsorbed at the same top positions. 
For FeMn, the
Fe-Cu$_{\text{Fe}}$ distance is very close to the value for the isolated
atom (2.37~\AA{} vs. 2.38~\AA). However, it increases for longer chains,
with a value of 2.40~\AA{} for FeMn$_6$. The same behavior is observed
for the Mn atoms, again close to the single adsorbate for the case of FeMn
(2.50~\AA~vs. 2.47~\AA), and it increases for longer chains reaching 2.56~\AA~for
FeMn$_6$. Hence, as the number $n$ of Mn atoms increases, we find
that the intra-chain distances diminish, see Table~\ref{distance1}, while the
distance between the chain atom and the top Cu ones increases.  Both
behaviors can be rationalized as stress built up as the chain increases its
size. Indeed, the elongation of these distances can be easily identify with
some destabilization of the chains as length is increased.
 
%\begin{table}[t]
%\centering
%\begin{ruledtabular}
%\begin{tabular}{lccccccc}
% & $E_{at}$ & $E_F$ & $\Delta_nE_F$  \\
% & (eV/TM atom) & (eV) & (eV)  \\
%        \hline
%        Fe         & 2.947 &  &   \\
%        Mn         & 1.821 &  &   \\
%        FeMn       & 2.659 & 0.550 & 0.550  \\
%        FeMn$_2$   & 2.522 & 0.977 & 0.427  \\
%        FeMn$_3$   & 2.443 & 1.364 & 0.387  \\
%        FeMn$_4$   & 2.399 & 1.763 & 0.399  \\
%        FeMn$_5$   & 2.367 & 2.149 & 0.386  \\
%        FeMn$_6$   & 2.339 & 2.500 & 0.351  \\
%\end{tabular}
%\end{ruledtabular}
%\caption[] {\label{energy} Values of atomization and formation energies, as defined in the text}
%\end{table}

\begin{figure}[ht]
\includegraphics[width=1.0\columnwidth]{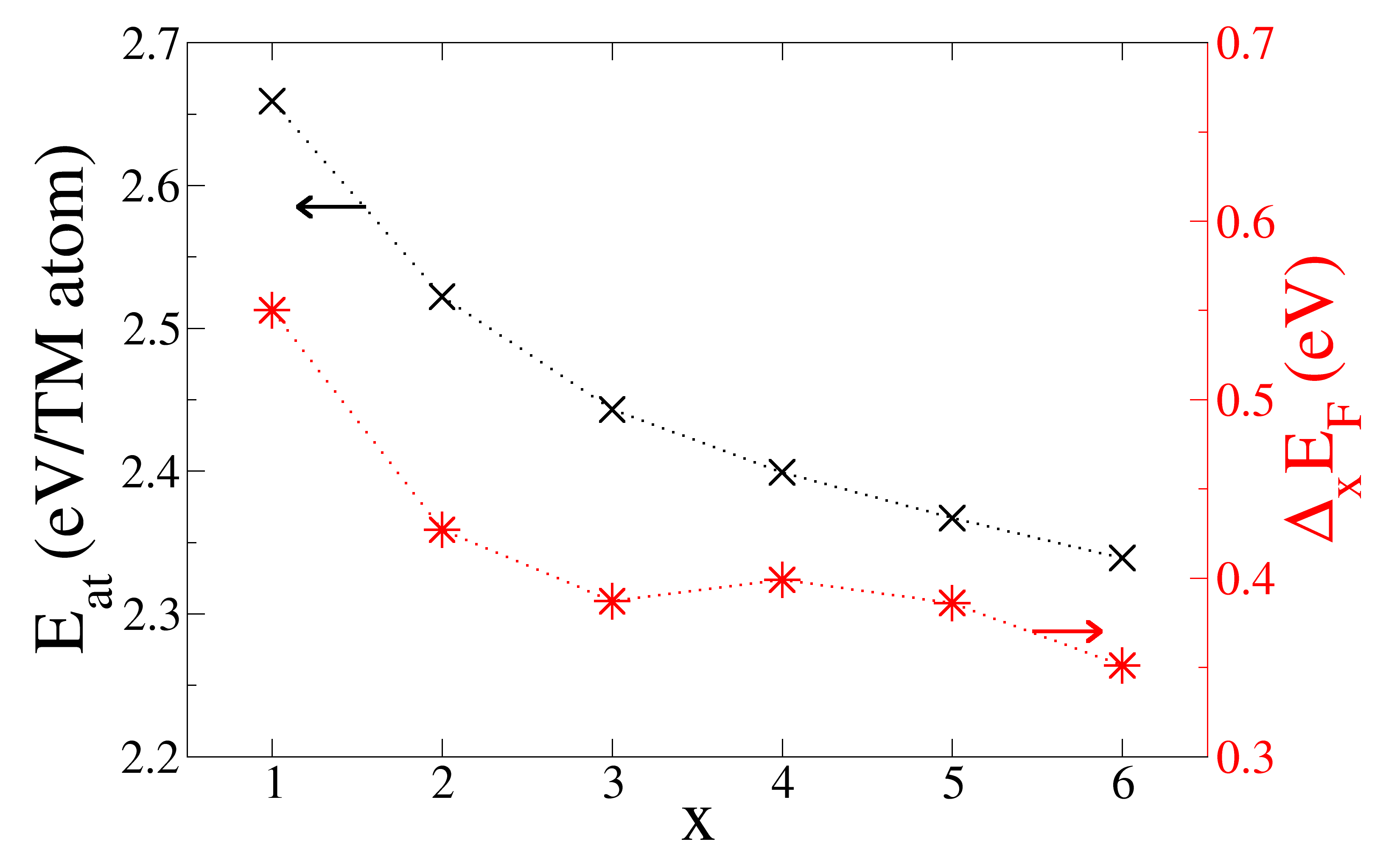}
\caption{\label{fig_energy} Atomization energy, $E_{at}$ from Eq.~(\ref{Eat}), (black x's) and
the gain in energy by adding one more Mn atom the chain, $\Delta_x E_F$ from
Eq.~(\ref{ef}), (red stars). As we can see, atomization of the chain
becomes less difficult as the chain becomes longer, and correspondingly, adding one
more atom does not reduce the total energy as much for longer chains. These values reveal an increasing destabilization
of the chain as it size increases.
The dotted lines are a guide for the eye.}
\end{figure}

To check for the stability of the chains we have studied the chain's
energetics. First, we start by analyzing
the atomization energy per TM atom, $E_{at}[FeMn_x]$, defined from
\begin{eqnarray} 
\label{Eat}
E_{at}[FeMn_x]&=& \\
-\frac{1}{x+1}[E[FeMn_x]&-&
E[Fe]-xE[Mn]-E_x[Cu_2N]], \nonumber
\end{eqnarray} 
where E[FeMn$_n$] is the total energy of the system;
E[Fe] (E[Mn]) is the energy of a gas-phase Fe (Mn) atom; and
E$_x$[Cu$_{2}N$] is the energy of the [3 $\times$ ($x$+3)] unit cell of the
bare surface. The atomization energy per TM atom is a measure of how much
{average energy per atom} one needs to give to the adsorbed chain to separate it
in its constituent Fe and Mn gas-phase atoms and the pristine Cu$_2$N/Cu(100) substrate.
Figure \ref{fig_energy} shows
the atomization energy per TM atom as a function
of the number of Mn atoms, $x$.
The atomization energy per TM atom tends to decrease
as the chain size increases,  
{ implying that long chains become energetically less favorable.}

We can extract more information by looking at another quantity that 
reflects better the experimental procedure to construct the chain. Let us
remember that the chains are constructed by STM manipulation. First,
Fe and Mn atoms are deposited on the surface, and then they are moved
with the STM tip to form the desired structure. To account for this
procedure, we can define a formation energy where the reference of energy is the
one of the
deposited TM atom, and then to study the gain in energy by adding
one more atom to the chain.
The definition that we used for the formation
energy is given by
\begin{equation}
\label{ef}
E_F[FeMn_x]=-[E_{at}[FeMn_x]-E_{at}[Fe]-xE_{at}[Mn]],
\end{equation}
where we use the atomization energies defined in equation~\ref{Eat}. From $E_F$ we can define
the gain in energy by adding one more TM atom, $\Delta_x E_F$\cite{robles_2010}, as
\begin{eqnarray}
\Delta_xE_F=E_F[FeMn_x]-E_F[FeMn_{x-1}]=\nonumber\\
-[E_{at}[FeMn_x]-E_{at}[FeMn_{x-1}]-E_{at}[Mn]].
\end{eqnarray}
This quantity, $\Delta_xE_F$, defines the energy gained by adding a Mn atom
to an existing chain starting from an Fe atom, which mimics the
experimental procedure to build the system.  The calculated values
of $\Delta_x E_F$ are always positive indicating that building
the chains is an energetically favorable process. The energy gain
is modest ($\approx0.4~eV$), and decreases by 37\% from FeMn to FeMn$_6$
(Fig.~\ref{fig_energy}). 
{Experimentally, the longest chains contained
9 Mn atoms.~\cite{choi_2014} Our results seem to suggest that longer chains will
be difficult to form due to the accumulated stress
imposed by the surface lattice constant and the Fe--Mn distance.}

\begin{table*}[t]
\centering
\begin{ruledtabular}
\begin{tabular}{lcccccccc}
& $\mu_T$ & $\mu$[Fe] & $\mu$[Mn$_1$] & $\mu$[Mn$_2$] & $\mu$[Mn$_3$] & $\mu$[Mn$_4$] & $\mu$[Mn$_5$]
& $\mu$[Mn$_6$]\\
        \hline
        Fe       &  3.82 & 3.35 &  &  &  &  &  & \\		
        Mn       &  4.97 &      &  4.80 &  &  &  &  & \\	
		FeMn     & -1.30 & 3.16 & -4.65 &  &  &  &  & \\				
		FeMn$_2$ &  3.68 & 3.17 & -4.55 & 4.71 &  &  &  & \\
		FeMn$_3$ & -1.43 & 3.19 & -4.56 & 4.56 & -4.73 &  &  & \\
		FeMn$_4$ &  3.78 & 3.22 & -4.56 & 4.57 & -4.57 & 4.75 &  & \\			
		FeMn$_5$ & -1.36 & 3.23 & -4.56 & 4.58 & -4.58 & 4.58 & -4.77 & \\		
		FeMn$_6$ &  3.86 & 3.23 & -4.57 & 4.58 & -4.58 & 4.59 & -4.58 & 4.79 \\
\end{tabular}
\end{ruledtabular}
\caption[] {\label{magnetic} Magnetic moments (in $\mu_B$) of the TM atoms in the chains. Values for
isolated Fe and Mn atoms on the surface are included for comparison. The values
have been obtained by performing a Bader analysis.\cite{tang_bader_2009} The total
magnetic moment $\mu_T$ of the calculation cell also includes contributions from
nearby copper atoms that spin polarize, hence $\mu_T$ slighlty differs
from the sum of the magnetic moments of Fe and Mn.}
\end{table*}

\subsection{Electronic and magnetic properties and charge transfer}

The magnetic properties are one of the main motivations for the study
of these systems. They largely stem from the electronic structure of
the atomic chains on the surface. An analysis of the electronic
structure by projection onto atomic orbitals have been shown
in Ref.~[\onlinecite{choi_2014}]. The magnetic properties were
shown to be given by the population of Fe and Mn $d$ orbitals.
In the case of Fe the d$_z^2$ orbital was basically filled {(where $z$ is
the direction perpendicular to the surface \textit{in the present case})}, not
contributing to the atomic magnetic moment. The Mn atoms
show a larger spin polarization due to the complete polarization
of its $d$-shell after inclusion in the FeMn$_x$ chain. The exchange splitting
of $d$-bands is the smallest for Fe, but still it is roughly
$5$ eV, showing how robust magnetism is in these chains. The
electronic structure near the Fermi energy is largely due to
the Cu and N states and with negligible magnetic polarization.

It is interesting to inspect the total magnetic
moment of the chains shown in Table~\ref{magnetic}. We can see how the
magnetic couplings between Fe-Mn and Mn-Mn atoms are antiferromagnetic
(AF) in the most stable configuration. In addition, the magnetic moment of
atomic Fe on the surface is 3.35$\mu_B$ (formally S=2), while the value
for Mn is 4.80$\mu_B$ (formally S=$5\over2$). Therefore, formally when
adding Mn atoms to a Fe one we would get systems with S=$1\over2$ for an
odd number of Mn atoms, $x$, and S=$2$ for an even number. In the computed
results for the total magnetic moment in the cell ($\mu_T$) we observe
the expected even--odd behavior, with values of 1.30--1.43 Bohr magnetons, $\mu_B$, for 
$x$ odd, and 3.68--3.86$\mu_B$ for the $x$ even. 
Values of 1.30--1.43$\mu_B$ correspond well with
the simple result S=$1\over2$. The difference from 1$\mu_B$ 
has two different sources that are present in these calculations.
On the one hand, the TM atoms present 
 fractional occupancies that lead to non-integer multiples of $\mu_B$.
On the other hand, these calculations
are  mean-field approximations to the difficult solutions
of  correlated AF ground states. As a 
consequence, the mean-field solution averages over
the possible atomic magnetic moments found in the 
multiple spin configurations of the correct AF ground state.

\begin{figure}[ht]
\includegraphics[width=\columnwidth]{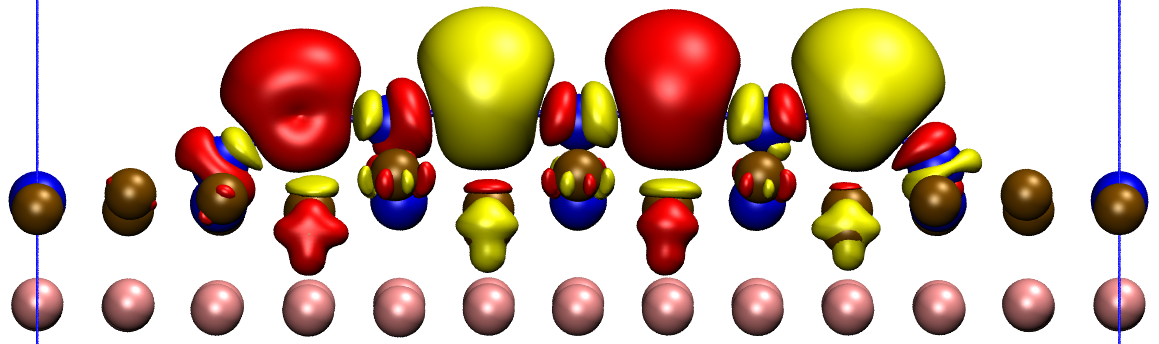}
\caption{\label{spin_density} Spin density of the FeMn$_3$ chain. Red (yellow) indicates majority 
(minority) spin areas. The chosen isovalue is 0.01~eV/\AA$^3$. The color code of the atoms is the 
same of Fig.~\ref{trimer}.}
\end{figure}

The AF coupling between the TM atoms is mediated
by the N atoms $via$ a superexchange mechanism as 
previously shown for  Mn~\cite{rudenko_2009} and Fe chains.~\cite{nicklas_2011} 
To illustrate the superexchange interaction,  we show
the spin-difference-density of the FeMn$_3$ chain in Fig.~\ref{spin_density}. The
spin polarization of the intercalated N atoms adopt the expected form
for superexchange, with induced spin polarization within the atom,
but net spin close to zero. Indeed, the bigger values for the induced magnetic
moments of N atoms are very small, for example 0.11$\mu_B$ for N$_1$ and 0.08$\mu_B$ for N$_2$.
Superexchange also leads to a change in the atomic angles: 
the Fe-N-Mn and Mn-N-Mn angles tend to approach 180$^\circ$ maximizing the 
AF interactions.~\cite{rudenko_2009,nicklas_2011}

The magnetic moments tend to increase with increasing chain size,
which is consistent with the previously mentioned progressive separation
of the chain from the surface. The magnetic moments of the Mn atoms within
a given chain are quite constant, with the exception of the final atom,
which is $\approx$4\% bigger. The reason is the lower coordination of
the edge atom, reflected in the slightly longer distance with the Cu
atom below and the longer distance with the adjacent N atoms.

The analysis of charge transfer in the system yields that each TM atom in the chain
loses around one $s$ electron to form the bond with N atoms in the
chain and the Cu atom underneath the TM atom. The $d$ charge is similar to the atomic
case, i.e., there are 6 electrons in the $d$ manifold for Fe, and 5 for Mn, 
in agreement with the computed magnetic moments.

\subsection{Exchange coupling constants}
\label{U-J}

\begin{table*}[t]
\centering
\begin{ruledtabular}
\begin{tabular}{lcccc cccccc}
& \multicolumn{4}{c}{1st neighbors} & \multicolumn{3}{c}{2nd neighbors} & \multicolumn{2}{c}{3rd neighbors}
& \multicolumn{1}{c}{4th} \\
& $J_{12}$ & $J_{23}$ & $J_{34}$ & $J_{45}$ & $J_{13}$ & $J_{24}$ & $J_{35}$ & $J_{14}$ & $J_{25}$ & $J_{15}$ \\
\hline
        \hline
                FeMn    & -10.25 &       &       &       &      &      &      &       &       & \\
                FeMn$_2$ & -6.51 & -2.09 &       &       & 0.97 &      &      &       &       & \\
                FeMn$_3$ & -7.30 & -1.97 & -1.97 &       & 1.14 & 0.32 &      &  0.10 &       & \\
                FeMn$_4$ & -7.96 & -2.69 & -2.10 & -2.16 & 0.73 & 0.05 & 0.27 & -0.36 & -0.01 & -0.02\\
\end{tabular}
\end{ruledtabular}
\caption[] {\label{jotas} Values for the exchange coupling constants (in meV) obtained by fitting Eq.~(\ref{heisenberg})
to the GGA+U total energies.}
\end{table*}

We have fitted our results to the Heisenberg Hamiltonian shown in
Eq.~(\ref{heisenberg}). In order to do so we have calculated
different spin solutions corresponding to  different magnetic
ordering among the TM atoms in the chain. In all calculations, we have
used the geometry of the most stable configuration, which is always
the AF case shown in Table~\ref{magnetic}. For the atomic values of $S$, in
Eq.~(\ref{heisenberg}), we had the choice of either using the formal
value (S=2 for Fe, S=$5\over2$ for Mn), or taking the computed value
from the DFT calculation since for a single atom $S_z$ is a good quantum
number. We have opted for the later option, which
gives values for the exchange coupling constants $J$'s which are between
15\% and 20\% higher. We can justify this choice because the DFT values
reflect the local magnetic moment that interacts via other atoms (superexchange)
with the local magnetic moment of the next neighbor in the calculation. 
The results are shown in Table.~\ref{jotas}. As
expected, $J$'s involving first neighbors are negative, which indicate an
AF coupling. Second neighbor $J$'s are positive, indicating a ferromagnetic
coupling which further stabilizes the global AF solution. Further order
J's are smaller, and its omission in the fit only implies a small error. 
 
The convergence with the number of coupled atoms considered in the
Heisenberg chain, Eq.~(\ref{heisenberg}), is very fast.
We have
done a systematic study for
FeMn$_4$ where different number of neighbors are included in solving
Eq.~(\ref{heisenberg}). Considering
just first neighbors introduces a maximum error of 0.2~meV. Considering
first and second neighbors the error is less than 0.1~meV. These calculations
show that the effective interactions in these spin chains are very short
range and first-neighbors truncation is indeed an accurate approximation.

%\begin{table*}[t]
%\centering
%\begin{ruledtabular}
%\begin{tabular}{lcccc cccccc}
%{\bf FeMn}$_4$ & \multicolumn{4}{c}{1st neighbors} & \multicolumn{3}{c}{2nd neighbors} & \multicolumn{2}{c}{3rd neighbors}
%& \multicolumn{1}{c}{4th} \\
%& $J_{12}$ & $J_{23}$ & $J_{34}$ & $J_{45}$ & $J_{13}$ & $J_{24}$ & $J_{35}$ & $J_{14}$ & $J_{25}$ & $J_{15}$ \\
%\hline
%        \hline
%                1st neighbors & -8.15 & -2.55 & -2.17 & -2.13 &      &      &      &       &       &       \\
%                2nd neighbors & -7.92 & -2.72 & -2.17 & -2.14 & 0.70 & 0.03 & 0.30 &       &       &       \\
%                3rd neighbors & -7.98 & -2.68 & -2.10 & -2.16 & 0.72 & 0.05 & 0.26 & -0.36 & -0.01 &       \\
%                All neighbors & -7.96 & -2.69 & -2.10 & -2.16 & 0.73 & 0.05 & 0.28 & -0.36 & -0.01 &  -0.03 \\
%\end{tabular}
%\end{ruledtabular}
%\caption[] {\label{neighbors} Values for the exchange coupling constants (in meV) for FeMn$_4$ considering
%different numbers of neighbors in Eq.~(\ref{heisenberg}).}
%\end{table*}

Analyzing the evaluated exchange couplings, $J$, we observe that the biggest value is obtained for
the first-neighbors interaction between Fe and Mn$_1$. Thus the AF coupling
between Fe and Mn is stronger than the one between Mn-Mn. 
FeMn is a special case since it involves two edge atoms, the
resulting AF coupling is the strongest one.
Curiously, the exchange coupling between Fe and Mn$_1$ ($J_{12}$) presents
a sharp drop for FeMn$_2$ to start increasing again for FeMn$_3$ and FeMn$_4$.
The coupling between Mn$_1$ and Mn$_2$ ($J_{23}$) presents
the same behavior, it shows a minimum for  FeMn$_3$.
Similar behavior is obtained for pure Mn chains
by Rudenko {\it et al.}.\cite{rudenko_2009}
These minima in the exchange couplings seem
to be independent from the atomic geometry, where
the behavior with chain length is monotonous, and with
the atomic magnetic moments. It is probably due to the
sudden appearance of an extra neighbor that symmetrizes
the interactions on the central atom of the chain.

\begin{figure}[ht]
\includegraphics[width=1.0\columnwidth]{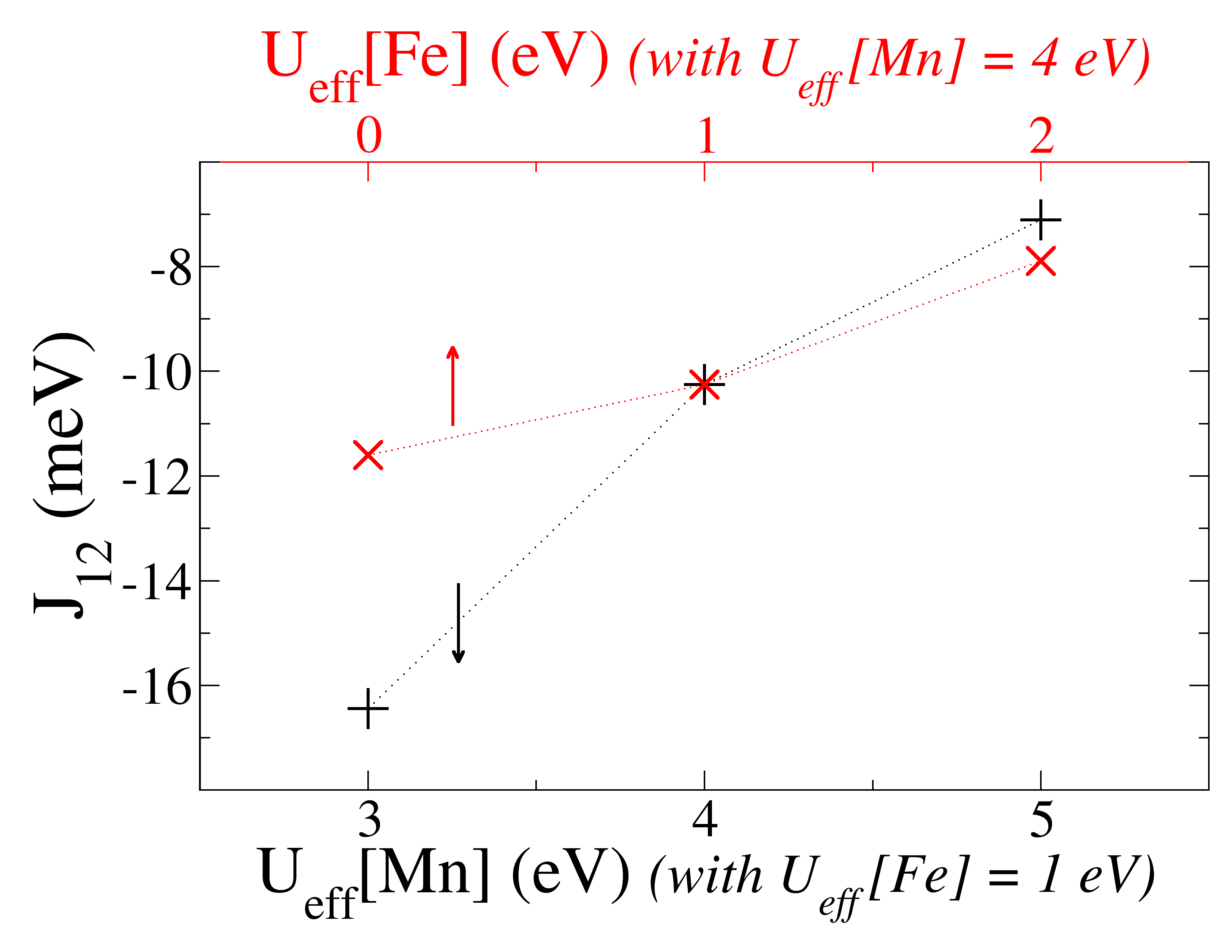}
\caption{\label{uj} Evolution of J$_{12}$ for FeMn with respect to the variation of $U_{\text{eff}}$.
We have changed one of the values of U while keeping constant the other one.}
\end{figure}
As Rudenko {\it et al.}~\cite{rudenko_2009} and Lin and
Jones~\cite{lin_2011} note, there is a dependence of the values of
the interatomic exchange coupling 
 on the chosen value of $U_{\text{eff}}=U-J$ for the GGA+U
approximation.
Figure~\ref{uj} shows
the dependence of $J_{12}$ for FeMn on $U_{\text{eff}}$.
We find that as the value of $U_{\text{eff}}$
decreases, $|J_{12}|$ increases. For the values of the
present study~\cite{lin_2011} ($U_{\text{eff}}$ [Fe] = 1~eV and  $U_{\text{eff}}$
[Mn] = 4~eV)  $J_{12}$ = -10.25~meV. If $U_{\text{eff}}$[Fe] is
reduced by 1~eV this value changes to -11.60~meV. When $U_{\text{eff}}$[Mn] is
reduced by 1~eV the effect is more pronounced, obtaining a value of -16.44~meV
(60\% increase in absolute value). 
These values give us an interval of exchange couplings that will be compared
with our experimental results in the following sections.

\subsection{Magnetic anisotropy}

\begin{table}[t]
\centering
\begin{ruledtabular}
\begin{tabular}{lcccc}
        MAE (meV)        & dir & $E_z-E_y$ & $E_x-E_y$ \\
        \hline
                Fe       & $y$ & 1.88 & 0.74  \\
                Mn       & $z$ & -0.13 & -0.01  \\
                FeMn     & $y$ & 1.51 & 0.75  \\
                FeMn$_2$ & $y$ & 1.46 & 0.48  \\
\end{tabular}
\end{ruledtabular}
\caption[] {\label{mae} Magnetic anisotropy energies (MAE) (in meV). We also show the easy axis directions, where $x$ is the direction in plane and perpendicular to the chain, $y$ along the chain, and $z$ perpendicular to the surface.}
\end{table} 

We have studied the magnetic anisotropy energy (MAE) of the system by
including SOC in our calculations. We have performed self-consistent
calculations in three directions: along the chain ($y$), perpendicular
to the chain in the plane ($x$), and perpendicular to the plane ($z$). We have
also fitted the  anisotropic spin Hamiltonian of Eq.~(\ref{ani}) {to
the singly adsorbed atoms using their computed magnetic moments} to obtain
values for $D$ and $E$ using the equations:
\begin{eqnarray}
D&=&\frac{2E_z-(E_x+E_y)}{S(2S-1)} \nonumber \\
E&=&\frac{E_x-E_y}{S(2S-1)},
\label{DE}
\end{eqnarray}
where $E_x$, $E_y$, $E_z$ are the magnetic anisotropy energies
when the spin, $S$, is aligned along the $x$, $y$ and $z$ directions, see
Table~\ref{mae}.

For
the supported atoms, Fe shows an easy axis along the chain, with a MAE of
1.88~meV. For Mn we get an out-of-plane easy axis, with a smaller MAE of
0.13~meV. This results qualitatively agree with the STM experimental data
of Hirjibehedin {\it et al},\cite{hirjibehedin_2007} although our values
for D and E are underestimated for Fe (D=-0.67 meV and E=0.29 meV using the DFT magnetic moment
to obtain $S$ in Eq.~(\ref{DE}))
 but in good agreement for Mn.  Shick
and co-workers~\cite{shick_2009}  evaluated within LDA the values for D
and E of Fe and Mn on Cu$_2$N/Cu (100). Their values for Mn agree with our
calculation, probably because the MAE of Mn is so small that the value
is within the error bar of our calculations. However, our results for
Fe are closer to the experimental values of D=-1.55 meV and E=0.31
meV. {Our calculations compared better with} the values by Shick {\it et al.}~\cite{shick_2009} 
{if the nominal magnetic moments are used (D=-0.44 meV and
E=0.19 meV using $S=2$ for
Fe. For Mn the values do not change because the DFT magnetic moment agrees
with the nominal one)}.  Using GGA+U, Barral et al~\cite{barral_2010}
obtained similar results to  Shick {\it et al.}~\cite{shick_2009}.
Recent calculations by Panda {\it et al.}~\cite{Panda2016}
yield similar values for Fe with LDA, and values
closer to the experiment using dynamical mean field theory (DMFT).

Taking into account that Mn and Fe have perpendicular easy axes,
one may wonder if SOC might induce a non-collinear alignment of the
magnetic moments in the chains. We have tested that possibility for
the shorter chains by performing an explicit non-collinear calculation
including SOC.\cite{hobbs_2000} Due to the small energies involved
in the calculation, the convergence of non-collinear solutions is very
challenging. We have been able to stabilize a non-collinear configuration
for FeMn where the magnetic moment of Fe is along the chain, while the
moment for Mn forms an angle of $27^{\circ}$ with the $y$ axis. However,
this solution is 3.79~meV higher in energy than the collinear solution.
For longer chains we have not been able to stabilize any non-collinear
solution.  Therefore, we have just considered collinear configurations
for the rest of our calculations.

The larger MAE of Fe forces the full-chain magnetic axis to align
with the Fe easy axis.  This is indeed seen in Table~\ref{mae}, where
the value for MAE decreases when increasing the chain length. This
reduction in MAE can be understood in terms of the addition of Mn atoms
that tend to align their magnetic easy axis perpendicular to the Fe one.
These results qualitatively agree with the results of anisotropic spin
Hamiltonians {where the total
magnetic anisotropy is approximated by adding the individual contributions of $D$ and $E$.}
This approximation does not capture the changes in geometry of the chain with
increasing lentgth. {Nevertheless, adding up the MAE contributions of each atom
leads to overestimations of MAE that are below the accuracy
of our DFT calculations.}

{The study of the magnetic anisotropy in these chains reveals their
composite structure. All FeMn$_x$ for odd-$x$ show a \textit{quasi}-spin of 1/2,
which should present zero anisotropy
if these chains were macrospins of spin 1/2, see. Eq.~(\ref{ani}). 
Our DFT calculations show sizeable anisotropies that underscore
the complexity of the magnetic states of these antiferromagnetic
structures.}

\subsection{Magnetic excitation energies}

\begin{figure}[ht]
\includegraphics[width=\columnwidth]{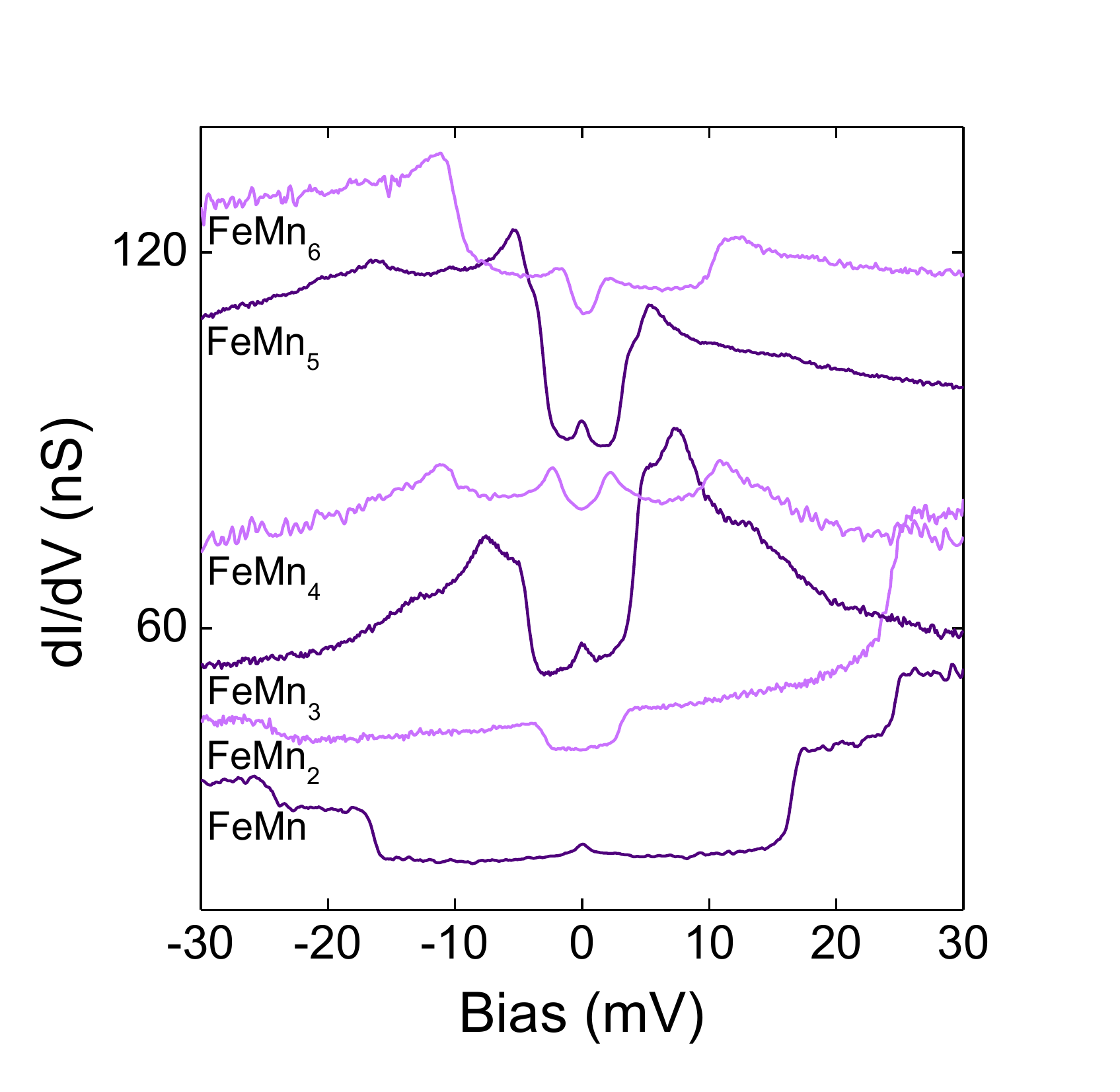}
\caption{\label{iets} 
{
Inelastic electron tunneling spectra (IETS) of different
FeMn$_x$ chains obtained by measuring the differential conductance with
the STM tip placed on a central Mn atom.}
}
\end{figure}

The magnetic structure of the chains can be obtained by studying the inelastic
electron tunneling spectra (IETS)~\cite{heinrich_2004} obtained with the STM. 
Figure~\ref{iets} shows the differential conductance obtained for the different
FeMn$_x$ chains. The features that appear in these spectra are due to magnetic
excitations, very similarly to the ones of Mn$_x$ chains shown in Ref.~[\onlinecite{hirjibehedin_2007}].
{Nevertheless,}
there are noticeable differences regarding both the peak at zero bias
for the odd-$x$ chains and the detailed structure of the steps. Indeed,
odd-$x$ chains are singlets in the case of Mn chains, here however, the 
ground states of odd-$x$ chains present 
a doublet ($S\approx \frac{1}{2}$) magnetic structure. 

We solve Eq.~(\ref{heisenberg}) and study
the different magnetic states for each chain using the
methodology of Ref.~[\onlinecite{Ternes_2015}]. The study of the ground state
is particularly important to understand the behavior of the Kondo
physics appearing in Fig.~\ref{iets}.\cite{choi_2014} Indeed, all odd-$x$ chains
display a Kondo zero-bias peak, while this peak is absent from the
chains with even $x$. Due to the Heisenberg correlations, the
experimental ground state is multi-configurational.
Solving Eq.~(\ref{heisenberg}) with the anisotropy terms, Eq.~(\ref{ani})
shows that many of these configurations are 
$S={1\over2}$ states, with a weight larger than 20\% in the total state. Hence,
a spin-flip process is possible at zero-energy cost, which explains
the appearance of Kondo peaks. Likewise, even-$x$ chains do 
not have degenerate ground states and Kondo physics is absent.  
The absence of degenerate ground states for even-$x$ is due 
to a total spin $S \approx 2$ where the large longitudinal MAE of
the Fe atom lifts the degeneracy of the ground state. The presence of
anisotropy prevents $S$ from being a good quantum number.\cite{choi_2014}

\begin{figure}[ht]
\includegraphics[width=0.8\columnwidth]{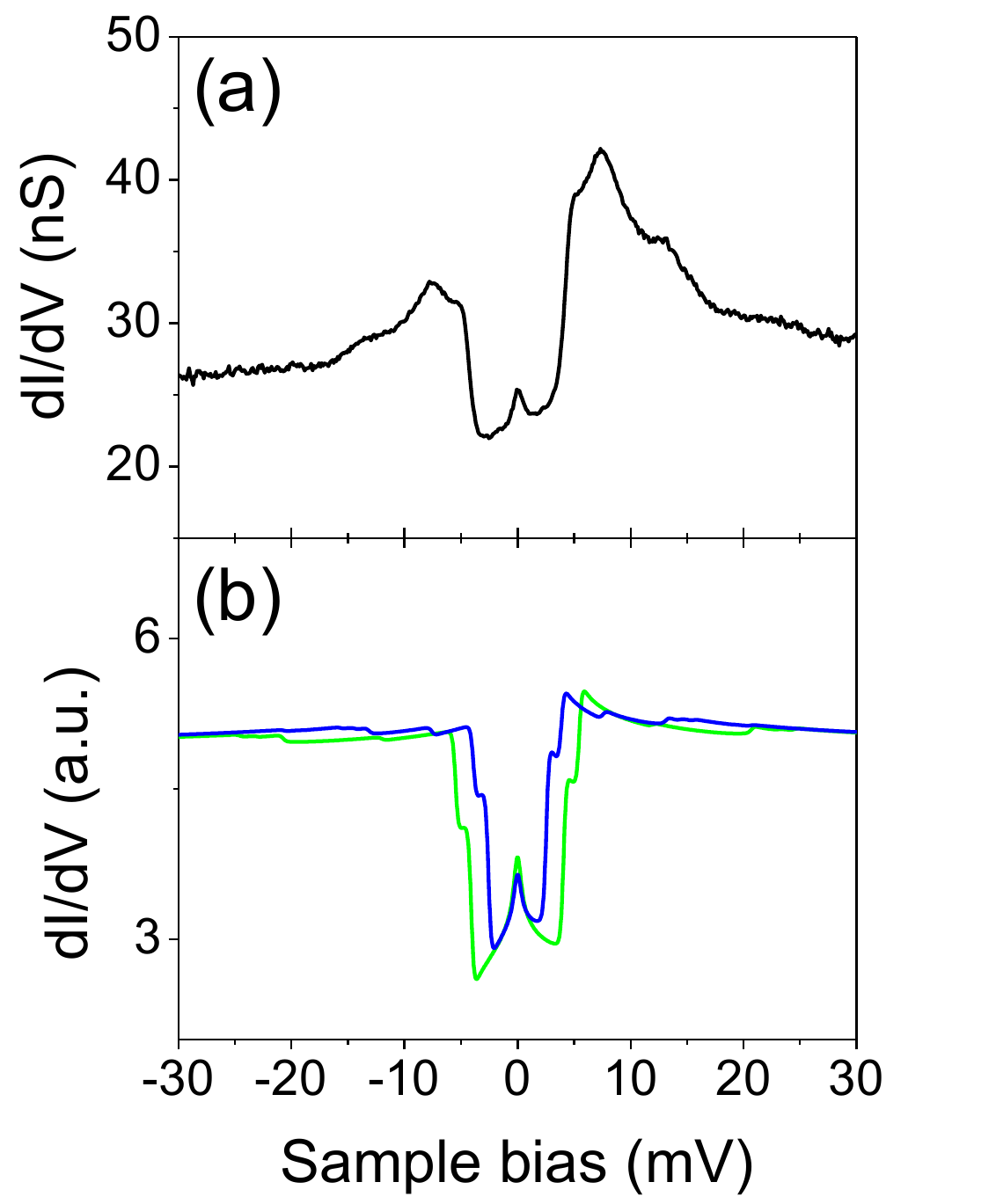}
\caption{\label{exp}
{
Inelastic electron tunneling spectra (IETS) of 
FeMn$_3$, $(a)$ experimental, $(b)$ 
two computed solution using our
calculated spin-chain parameters for $(U-J)_{\text{Mn}}=4$ eV
and $(U-J)_{\text{Fe}}=1$ eV (blue) or
scaling the computed exchange coupling
by a 1.6 factor as found for FeMn with $(U-J)_{\text{Mn}}=3$ eV
and $(U-J)_{\text{Fe}}=1$ eV (green), section~\ref{U-J}.
The STM tip is placed on the edge Mn atom.}
}
\end{figure}

The solutions of  Eq.~(\ref{heisenberg}) permit us to compare the computed
data for the chains with the experimental data.  Figure~\ref{exp}, $(a)$
shows the IETS for the FeMn$_3$ chain measured of the edge Mn atom,
and $(b)$ shows two calculations. The first one (blue) performed with
{the exchange couplings computed from}
our GGA+U calculations with an effective $U$ value for
Mn of 4 eV and for Fe of 1 eV, see section~\ref{U-J},
{and adding up the experimental atomic magnetic anisotropies.} The second one
(green) scales the exchange couplings by 1.6, using the scaling found
for FeMn when the effective $U$ value for Mn was reduced to 3 eV, and
keeping the corresponding value of Fe constant. We see that the energy
thresholds are in good agreement in the second case, and the solution
of Eq.~(\ref{heisenberg}) with the third-order perturbation method
of Ref.~[\onlinecite{Ternes_2015}] largely  reproduces the dynamical
phenomenology of the magnetic chain including higher-energy excitations
like the one at $\pm 10$ meV.

The asymmetry of the main inelastic thresholds
found in Fig.~\ref{exp}$(a)$ is treated
within the third-order perturbation method
by including a potential scattering term
in the Kondo scattering.\cite{Ternes_2015}
The effect of the potential scattering term is
to remove the electron-hole symmetry of the
excitation spectra of Eq.~(\ref{heisenberg}).
In order to fit the experimental spectra
we have used a $J_{\text{Kondo}} \;  \rho = -0.04$ (where
$J_{\text{Kondo}} $ is the Kondo exchange coupling
with electrons from the substrate and $\rho$ is
the density of states at the Fermi energy)
and a potential scattering term ${U_{\text{Kondo}}}/{J_{\text{Kondo}}}=-0.5$.
The results indicate that larger values should be used
to reproduce the experimental data, implying the
need to go beyond third-order perturbation theory
to treat Kondo scattering in FeMn$_3$.

\begin{figure*}[ht]
\includegraphics[width=1.7\columnwidth]{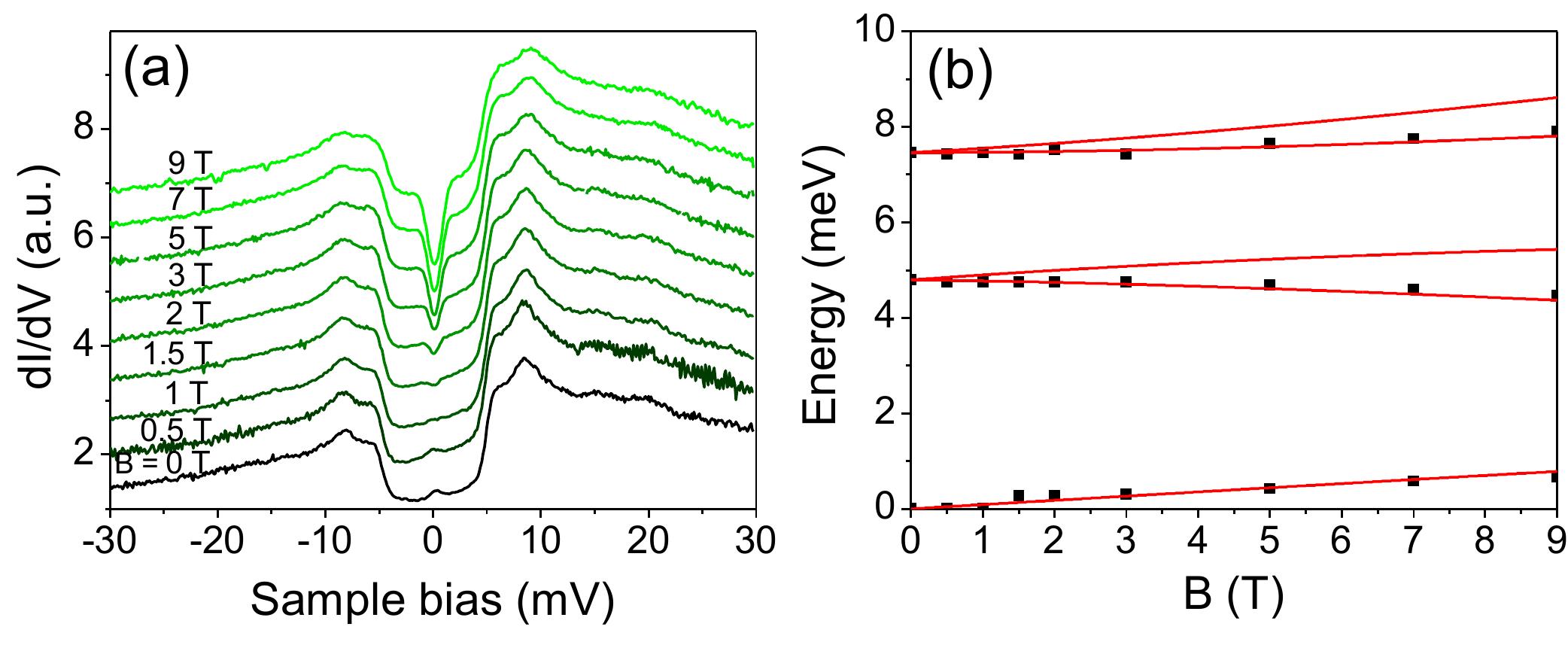}
\caption{\label{magnet} 
{
$(a)$ Inelastic
electron tunneling spectra (IETS) for FeMn$_3$ with a magnetic field
applied perpendicular to the surface increasing in 1-T steps until
9 T. $(b)$ The behavior of the different magnetic states (black dots) obtained from the
experimental figure follows the Zeeman trend expected for a Zeeman
term with the atomic  $g$-factors
{($g_{Fe}=2.1$
and $g_{Mn}=1.9$ from Ref.~[\onlinecite{hirjibehedin_2007}])}.
}
}
\end{figure*}

The magnetic behavior of
the inelastic thresholds 
is correctly reproduced by a  Zeeman shift.
This permits us to extract the value of
the gyromagnetic ratio $g$, Fig.~\ref{magnet}. 
For the present case we find that the atomic g's {($g_{Fe}=2.1$
and $g_{Mn}=1.9$ from Ref.~[\onlinecite{hirjibehedin_2007}])} are good
approximations to obtain the correct behavior of the magnetic global
states with external B, Fig.~\ref{magnet}.  To a large
extend, the atomic spin preserves its character, although very entangled
due to the sizable Heisenberg exchange interactions.

\section{Discussion and concluding remarks}

This work is a detailed experimental and theoretical account of the
electronic and magnetic properties of a heterogeneous type of magnetic
atomic chain (FeMn$_x$ with $x=1,6$) adsorbed on Cu$_2$N/Cu (100).
The chains are assembled by atom manipulation with an STM tip, and
stable configurations are found when the TM atoms (TM = Fe and Mn) sit
ontop of Cu atoms. Experimentally, it is difficult to assemble chains
with $x > 9$ {(we rarely succeeded going beyond $x=10$
creating a straight FeMn$_x$ chain)} and theoretically we see that stress builds up as the
chain increases size due to the imposed TM-TM distance by
the underlying Cu$_2$N/Cu (100) substrate. As the chain increases its
size, the TM atoms increase their mutual distance and also their
distance to the chain, energetically this is translated into a systematic
lowering of both the atomization energy and the energy gained by the chain
every time a new Mn atom is added.

Upon adsorption, Fe and Mn lose one of their $s$ electrons in the interaction
with the substrate, mainly to form the bond with the neighboring N atoms. There
is considerable distortion and hybridization of the d-electron structure but
their occupations remain the free-atom ones, leading to magnetic moment
values close to the gas phase.\cite{hirjibehedin_2007,lin_2011}

The experimental data involve the IETS of different chains with detailed
information on the excitation energies of the chains. The lower-energy
spectral features are due to magnetic excitations of the system as was
tested by their magnetic field dependence.  In parallel, the values of
the Heisenberg Hamiltonian can be obtained from DFT by evaluating the
energy of different spin arrangements of the chains. This approach gives
us a systematic insight on how the different TM atoms relate to each other
in the chain, which are mainly driven by antiferromagnetic superexchange
mediated by the non-magnetic N atoms of the surface.  We find that the
Fe-Mn couplings are systematically larger than the Mn-Mn ones, and that
beyond second neighbors neglection of the magnetic coupling is a very
good approximation. Indeed, first neighbors is a sufficient approximation
to obtain exchange couplings with an error of 0.2~meV.  In the present
choice of DFT+U calculations, the computed magnetic exchanges lead to
excitation energies smaller than the experimental ones. 
Unfortunately, there is not a unique way of determining
the value of the Hubbard $U$ for
the calculations. Our systematic study of the values show that a change
of 1 eV in the value of the Mn $U$ leads to a $\sim$ 60\% change in the value
of the evaluated couplings yielding good agreement with the experiment.

The TM atoms are subjected to magnetic anisotropies on this surface. Our
calculations show that the MAE of the full chain is not just the sum
of the MAE's of each TM atom. Nevertheless, the smallness of Mn MAE
renders this approximation acceptable.  Despite their different easy
axis, the very large MAE of the Fe dominates and the Mn spins orient
along the chain following the Fe easy axis. This leads to a collinear
solution of the initially non-collinear problem. We have not found any
spin canting or frustration although we cannot rule it out for longer
chains.  {Moreover, the study of the magnetic anisotropy in these
chains reveals their composite structure. All FeMn$_x$ for odd-$x$ show
a \textit{quasi}-spin of 1/2.  If these chains were spins 1/2, their
anisotropy would be stricly zero.  However, our calculations show that
they have sizeable anisotropies in agreement with the complexity of the
magnetic states of these antiferromagnetic structures.}

This combined experimental and theoretical work
gives us direct insight into  the
different  electronic, geometric and magnetic properties
 of these heterogeneous chains. In particular, we
have given an account for the
appearance of Kondo peaks and the antiferromagnetic
character of these chains, their
magnetic anisotropy that permits us to rule out a macrospin behavior,
as well as the accumulated stress that limits the length
of the chains.

\acknowledgments
DJC acknowledges the European Union for support under the
H2020-MSCA-IF-2014 Marie-Curie Individual Fellowship programme proposal
number 654469 and a previous postdoctoral fellowhship from the Alexander von Humboldt foundation.  DJC and SL acknowledge Edgar Weckert and Helmut Dosch
(Deutsches Elektronen-Synchrotron, Hamburg, Germany) for providing
high-stability lab space.  NL acknowledges financial support from Spanish
MINECO (Grant No. MAT2015-66888-C3-2-R).
ICN2 acknowledges support from the
Severo Ochoa Program (MINECO, Grant SEV-2013-0295).

\bibliography{femn}
\bibliographystyle{apsrev4-1}

\end{document}